\documentclass{article}
\usepackage{datetime}
\usepackage{amsmath}
\usepackage{amsthm}
\usepackage{amssymb}
\usepackage{bbm}
\usepackage{hyperref}
\usepackage{algorithm}
\usepackage{algpseudocode}
\usepackage{natbib}
\usepackage[bottom]{footmisc}
\usepackage{xcolor}
\usepackage{titling}
\usepackage{tikz}
\usepackage{graphicx}
\usepackage{hyperref}
\usepackage{enumitem}
\usepackage{booktabs}
\usepackage{tcolorbox}
\usepackage{listings}
\usepackage{geometry}
\usepackage{pgfplots}

\newcommand{\Z}{\mathbb{Z}}

\newcommand{\Zp}{\mathbb{Z}_p}
\newcommand{\Zpm}{\mathbb{Z}_{p-1}}
\newcommand{\rdmpf}{\operatorname{RDMPF}}

\newcommand{\enc}{\operatorname{enc}}

\newcommand{\keyd}[2]{\text{key}_{#1\text{-}#2}}

\newcommand{\HKDF}{\mathsf{HKDF}}
\newcommand{\HMAC}{\mathsf{HMAC}}
\newcommand{\SHA}{\mathsf{SHA3\mbox{-}256}}
\newcommand{\Kenc}{K_{\mathrm{enc}}}
\newcommand{\Kauth}{K_{\mathrm{auth}}}
\newcommand{\ctkem}{ct_{\mathrm{kem}}}
\newcommand{\Hint}{\mathsf{HINT}}
\newcommand{\Inner}{E_{\mathrm{inner}}}
\newcommand{\W}{\mathcal{W}}
\newcommand{\Id}{\mathsf{id}}
\newcommand{\Commit}{\mathsf{Commit}}
\newcommand{\Finalize}{\mathsf{Finalize}}
\newcommand{\Announce}{\mathsf{Announce}}
\newcommand{\DestructIntent}{\mathsf{DestructIntent}}
\newcommand{\DestructProof}{\mathsf{DestructProof}}
\newcommand{\AEAD}{\mathsf{AEAD}}
\newcommand{\IV}{\mathsf{IV}}
\newcommand{\Notice}{\mathsf{Noticeboard}}
\newcommand{\Trunc}{\mathrm{Trunc}}
\newcommand{\Tuple}{\mathsf{tuple}}
\newcommand{\ctx}{\mathsf{context}}

\newcommand{\Setup}{\mathsf{Setup}}
\newcommand{\EncL}{\mathsf{EncL}}
\newcommand{\EncR}{\mathsf{EncR}}
\newcommand{\KeyGen}{\mathsf{KeyGen}}
\newcommand{\Eval}{\mathsf{Eval}}

\newcommand{\sk}{\mathsf{sk}}

\usetikzlibrary{positioning,shapes,arrows.meta,calc,backgrounds,fit,decorations.pathreplacing}
\pgfplotsset{compat=1.17}
\theoremstyle{plain}

\newtheorem{theorem}{Theorem}[section]
\newtheorem{lemma}[theorem]{Lemma}

\theoremstyle{definition}
\newtheorem{definition}[theorem]{Definition}
\theoremstyle{remark}
\newtheorem{remark}[theorem]{Remark}

\definecolor{customLink}{HTML}{0077B6}

\setcitestyle{authoryear,round}

\title{A Privacy Protocol Using Ephemeral Intermediaries\\
and a Rank-Deficient Matrix Power Function (RDMPF)\\
\vspace*{1em}
\normalsize An Application on the Internet Computer (ICP)}
\author{\vspace{-1em} Eduardo Salazar \\ \\ Nebula Technology Lab \\ \color{customLink}{\small{\href{mailto:eduardo.salazar@nebulalab.ae}{eduardo.salazar@nebulalab.ae}}}}
\date{\small{December 2025}}

\newenvironment{cogent}{\fontfamily{lmss}\selectfont}{\par}

\begin{document}

\maketitle

\begin{abstract}
This paper presents a private transfer architecture for the Internet Computer (ICP) that decouples deposit and retrieval through two short-lived intermediaries,
with sealed storage and attested teardown by an ephemeral witness. The protocol uses a non-interactive RDMPF-based encapsulation to derive per-transfer transport 
keys. A public notice hint is computed from the capsule to enable discovery without fingerprinting the recipient’s key. Retrieval is authorized by a short proof of 
decapsulation that reveals no identities. All transaction intermediaries are ephemeral and issue certified destruction intents and proofs, allowing a noticeboard to 
publish auditable finalization records. The design provides sender identity privacy with respect to the recipient, content confidentiality against intermediaries, 
forward secrecy for transport keys after staged destruction, verifiable liveness and finality. We formalize the basic interfaces, provide the security arguments for 
encapsulation correctness, hint privacy, authorization soundness and timeout reclaim.

In terms of implementation, it has been recently brought into production on the ICP under the name ICPP. It has been subject to exhaustive testing and incorporates 
several enhancements, focusing on the operational possibilities offered by ICP’s technology. This work hence serves as a broad reference for the protocol now live, 
which is publicly accessible at \href{https://icpp.tech}{https://icpp.tech}. 

\end{abstract}

\tableofcontents
\newpage


\section{Introduction}

\subsection{Overview}\label{sec:overview}
This paper presents an implementation of a \textbf{dual-intermediary dead-drop architecture} that cleanly separates deposit from retrieval, and provides witnessed 
teardown. In the discussion, we will resort to the usual characters Alice and Bob. The task for Alice (sender) is to transfer something to Bob (recipient) such that
the process remains opaque to external third parties. Specifically concerning implementation on the ICP, instances (computational units) are known as canisters, which 
are smart contracts that bundle both code and their persistent state. We will use the terms interchangeably.  

\begin{itemize}[leftmargin=1.4em]
  \item \textbf{Deposit intermediary $I_1$}\\
        Alice (via an ephemeral principal) submits a sealed package $\{\Hint,\ctkem,\Inner\}$ to an intermediary $I_1$. $I_1$ writes the package to storage instances
        $C_1,\ldots,C_n$, posts an $\Announce$ record with $\Hint$ on the certified $\Notice$, and then is destroyed after issuing a $\DestructIntent$ to its parent
        (or Factory) which is attested by a witness $\W$. Ephemeral instances are blackholed, hence they cannot self-destroy. 

  \item \textbf{Storage $C_i$ (sealed, dumb)}\\ 
        Each $C_i$ stores exactly the opaque tuple $\{\Hint,\ctkem,\Inner\}$. It never interacts with Alice or Bob directly. Upon being served at retrieval, 
        the $C_i$ instances are destroyed, with $\W$ recording the corresponding $\DestructProof(C_i)$.

  \item \textbf{Discovery on $\Notice$}\\ 
        Bob scans the $\Notice$ for his short public \textbf{hint} $\Hint=\SHA(\ctkem)$ and obtains the rendezvous token for $I_2$.

  \item \textbf{Retrieval intermediary $I_2$}\\
        Bob contacts $I_2$ and proves possession of the non-interactive RDMPF decapsulation by submitting $h=\SHA(K_{\mathrm{auth}})$, which $I_2$ verifies against 
        the value embedded in $\Inner$. $I_2$ gathers the stored tuple from a quorum of $C_i$, delivers $\Inner$ to Bob, then is destroyed via Factory after issuing 
        $\DestructIntent$. $\W$ records $\DestructProof(I_2)$.

  \item \textbf{Witness $\W$ and Finalization}\\ 
        $\W$ certifies storage commits and all staged destructions ($I_1$, served $C_i$, $I_2$) and finally publishes a $\Finalize$ record on the $\Notice$, then is 
        destroyed via Factory.
\end{itemize}

\noindent\textbf{Cryptographic core (non-interactive) --} Alice forms $\ctkem=(epk_S,\mathsf{nonce},\mathsf{tag})$ where
\[
\begin{aligned}
Z &\leftarrow \mathrm{RDMPF\_Key}\big((P_S,Q_S),(P_R,Q_R)\big),\\
(\Kenc,\Kauth) &\leftarrow \HKDF\!\big(Z;\ \textsf{info}=\text{"rdmpf-kem"},\ \textsf{salt}=\mathsf{nonce}\big),\\
\mathsf{tag} &\leftarrow \HMAC_{\Kauth}(\ctx).
\end{aligned}
\]
\noindent sets $\Hint=\SHA(\ctkem)$, and encrypts the payload as $\Inner=\AEAD_{\Kenc}(\textsf{payload})$. $K_{\mathrm{auth}}$ is used only 
once at $I_2$ to prove decapsulation; no instance ever learns RDMPF secrets.

\subsection{Security Properties}\label{sec:sec-props}
The protocol provides the following privacy and integrity properties under the assumptions in Section~\ref{sec:op-flow}.

\begin{enumerate}[leftmargin=1.6em]
  \item \textbf{Sender-identity privacy}\\ 
      Bob never learns Alice’s identity or a stable sender handle. The only sender material Bob may see is the one-time $epk_S$ (if included inside $\Inner$).
  \item \textbf{Content confidentiality vs. intermediaries}\\
       $I_1$, $I_2$, and $C_i$ only ever handle the opaque tuple $\{\Hint,\ctkem,\Inner\}$; the payload is protected by $\AEAD_{\Kenc}$.
  \item \textbf{Path unlinkability}\\ 
       No single ephemeral instance can link Alice to Bob: $I_1$ observes only the deposit, $I_2$ observes only the retrieval, and $C_i$ see neither party directly.
  \item \textbf{Forward secrecy on teardown}\\ 
      Transport keys $(\Kenc,\Kauth)$ are derived from a one-shot RDMPF secret $Z$ and become unrecoverable once the ephemeral state is destroyed and ephemeral 
      secrets are discarded; $\W$ attests staged destruction.
  \item \textbf{Recipient-private discovery}\\ 
      The public hint $\Hint=\SHA(\ctkem)$ leaks no information about Bob's public key beyond brute-force over 256 bits.
  \item \textbf{Security assumption}\\
      Security reduces to the hardness assumptions underlying RDMPF and standard primitives (HKDF/HMAC/SHA3/AEAD), not to any canister-local key rotation.
\end{enumerate}

\subsection{ICP-Specific Limitations and Threat Model}\label{sec:icp-threat}

\noindent\textbf{a.\ Privacy Guarantees (What is Hidden)}
\begin{itemize}[leftmargin=1.6em]
  \item \textbf{Transaction content}\\
      Opaque to all intermediaries. Only $\Inner=\AEAD_{\Kenc}(\textsf{payload})$ is handled in transit.
  \item \textbf{Sender identity from recipient}\\
      Bob cannot determine who sent the transfer, and Alice uses an ephemeral principal and one-time $epk_S$.
  \item \textbf{Path linkage}\\
      External observers and any single ephemeral instance cannot link deposit (via $I_1$) to retrieval (via $I_2$).
  \item \textbf{Recipient identity from intermediaries}\\
      $C_i$ never see Bob, hence $I_2$ authenticates a decapsulation proof $h=\SHA(K_{\mathrm{auth}})$ without revealing Bob’s long-term identity.
\end{itemize}

\noindent\textbf{b.\ ICP Exposure Model and Mitigations}

\noindent ICP's execution model exposes certain metadata by default. The ICPP protocol systematically neutralizes each exposure.

\begin{itemize}[leftmargin=1.6em]
  \item \textbf{Caller principals}\\
        Each canister call reveals its caller to that callee. \emph{Mitigation:} Alice and Bob use ephemeral principals; the $C_i$ canisters are never called by 
        either party directly. $I_1$ sees only an ephemeral sender principal (unlinkable to Alice's long-term identity), while $I_2$ sees only an ephemeral recipient 
        principal.
  \item \textbf{Call-graph metadata}\\
        Consensus records that calls occurred (not payloads). \emph{Mitigation:} The observable call graph is $I_1 \to C_i$ and $I_2 \to C_i$, which are 
        infrastructure-to-infrastructure calls that reveal topology but not endpoints. There are no Alice $\to C_i$ or $C_i \to$ Bob calls.
  \item \textbf{Ledger transfer correlation}\\
        ICP ledger transfers are public. \emph{Mitigation:} Deposits are held in functional subaccounts derived from $\mathsf{deposit\_id}$, not user principals. 
        Upon sealing, funds move to a shared mixer pool via randomized chunking, breaking amount-based and timing-based correlation. Subset-sum reconstruction is 
        computationally infeasible.
  \item \textbf{Canister state inspection}\\
        Canister memory could in principle be inspected. \emph{Mitigation:} All secrets ($\Kenc$, $\Kauth$, and RDMPF intermediates) exist only in transient RAM during 
        execution and are explicitly zeroed before destruction. Stable memory is never used for sensitive data. Post-destruction, canister memory is deallocated by the 
        subnet. No data at rest survives.
  \item \textbf{Subnet placement}\\
        Exact subnet assignment is platform-controlled. \emph{Mitigation:} Factory requests pairwise-distinct subnets for $I_1$, $I_2$, $C_i$, and $\W$. However, actual 
        placement cannot be guaranteed. Placement is recorded in certified state, and clients may enforce acceptance policies. Residual co-location risk is acknowledged 
        but minimized.
\end{itemize}

\noindent After destruction, the only persistent artifacts are the $\Notice$ records ($\Announce$ and $\Finalize$), which contain only $\Hint$ (a 256-bit hash), 
destruction proofs, and no other identifying information.\\

\noindent\textbf{c.\ Threat Model}
\begin{itemize}[leftmargin=1.6em]
  \item \textbf{Honest-but-curious intermediaries}\\
        $I_1$, $I_2$, $C_i$ may observe metadata but cannot break AEAD/HKDF/HMAC/SHA3 or RDMPF.
  \item \textbf{Network/global observer}\\
        Sees call graphs and timing but not contents. Discovery relies only on $\Hint$, not on recipient key fingerprints.
  \item \textbf{Compromised component(s)}\\
        Any \textbf{single} compromised ephemeral component cannot link Alice to Bob. Collusion of \textbf{both} $I_1$ and $I_2$ (or broad network compromise) is out 
        of scope and would defeat path unlinkability.
\end{itemize}

\noindent\textbf{d.\ Comparison to Other Privacy Systems}\\

\noindent It is widely acknowledged that popular privacy systems face \emph{infrastructure collusion} as a common vulnerability. We briefly compare below how this 
manifests across other privacy-oriented designs, to place our protocol in context.

\begin{itemize}[leftmargin=1.6em]
  \item \textbf{Monero --}
        Ring signatures hide the true input among decoys, but nodes can log transactions, IP addresses, and timing. A Sybil adversary controlling sufficient 
        nodes can apply traffic analysis to degrade unlinkability. Mitigations (Dandelion++, Tor integration) reduce but do not eliminate this vector.
  \item \textbf{Zcash --} 
        Zero-knowledge proofs provide strong on-chain privacy, but network-level metadata (origin IP, timing, pool behavior) remains observable. Colluding validators 
        can correlate this metadata despite cryptographic unlinkability. Orchard's unified actions improve arity hiding but do not address network-layer exposure.
  \item \textbf{Tornado Cash --} 
        On-ledger privacy is cryptographically strong, but relayers observe deposit/withdrawal requests with IP addresses, and Ethereum nodes see all transactions. 
        Colluding infrastructure can link deposits to withdrawals via timing and metadata.
  \item \textbf{Tor --} 
        Onion routing hides content, but exit nodes see destinations and directory authorities see topology. Collusion between entry (guard) and exit nodes enables 
        correlation attacks, compounded by low-latency design that preserves timing fingerprints.
  \item \textbf{Signal --} End-to-end encryption protects content, but Signal's servers observe delivery metadata. An adversary monitoring both the ISP/backbone 
        and Signal's infrastructure can correlate sender-recipient pairs. Sealed Sender mitigates this by hiding the sender from Signal's servers.
\end{itemize}

\noindent By comparison, our protocol addresses infrastructure collusion through \emph{architectural separation}: $I_1$ sees only the deposit (ephemeral sender), $I_2$ 
sees only the retrieval (ephemeral recipient), and $C_i$ see neither party directly. No single ephemeral component can link sender to recipient. The \emph{Sealed Sender} 
approach (hiding sender identity from intermediaries) is analogous to Signal's design, applied here to the transfer rather than messaging context. Collusion of 
\emph{both} $I_1$ and $I_2$ would be required to break path unlinkability, a threat that is explicitly out of scope (refer to Section~\ref{sec:icp-threat}c).


\section{Operational Flow}\label{sec:op-flow}

\subsection{System Setup}\label{sec:op-setup}

\begin{tcolorbox}[colback=blue!10,colframe=blue!75!black,title=Setup Phase]
\begin{enumerate}
\item System publishes public parameters $\pi_{\text{pub}} = (p, \dim, \text{BaseX}, \text{BaseY}, W)$
\item Alice generates key material. For each transfer, she uses a one-time public key $(P_S,Q_S)$ associated to $epk_S$.
\item Bob generates keypair $\rightarrow$ Private $(\lambda_R, \omega_R)$ \,|\, Public $(P_R, Q_R)$
\item Public keys are published to a purely pseudonymous registry
\end{enumerate}
\end{tcolorbox}

\subsection{Phase 1 -- Ephemeral Infrastructure Creation}\label{sec:phase1}
\begin{tcolorbox}[colback=blue!10,colframe=blue!75!black,title=Spawn Phase]
\begin{enumerate}[leftmargin=1.6em]
  \item \textbf{Protocol call --} Alice requests: \textit{Send $A$ ICP to $pk_R=(P_R,Q_R)$} with policy and TTL.
  \item \textbf{Factory spawn --} Spawn fresh $I_1, I_2, C_1,\ldots,C_n, \mathcal{W}$ with pairwise-distinct subnets and $\mathsf{subnet}(\mathcal{W})$ distinct.
  \item \textbf{Ephemeral control --} Self-controlled canisters, blackholed controllers, and one-shot destruct timer ($\le$ TTL).
  \item \textbf{Initialization --} Install verified WebAssembly (Wasm); set quorum ($t$-of-$n$ or replication); empty storage at each $C_i$. \emph{No RDMPF keys on 
        canisters.}
  \item \textbf{Identifiers \& routing --} Factory returns ids+subnets. Provision: $I_1\gets\{C_i\}+\Notice$; $I_2\gets\{C_i\}+\mathsf{id}(\mathcal{W})$;
        $C_i\gets\mathsf{id}(\mathcal{W})$; $\mathcal{W}$ records ids+subnets+Factory tx id as certified $\mathsf{context}$.
  \item \textbf{CSRN generation --} Each $C_i$ samples a 32-byte Canister-Signed Receipt Nonce for binding and mutual confirmation (Section~\ref{sec:csrn}).
  \item \textbf{Economics --} Alice prepays fees covering spawn/storage/ops/teardown with a retry margin.
\end{enumerate}
\end{tcolorbox}

\vspace{1em}
\noindent\underline{Initial knowledge}\\

\noindent $I_1$: $\{C_i\}$, $\Notice$. $I_2$: $\{C_i\}$, $\mathsf{id}(\mathcal{W})$. $C_i$: $\mathsf{id}(\mathcal{W})$. $\mathcal{W}$: ids, subnets, 
$\mathsf{context}$. No canister holds identities or RDMPF secrets.

\subsection{Phase 2 -- Alice Deposits via $I_1$}\label{sec:phase2}
\paragraph{RDMPF non-interactive encapsulation (interface) --}
Let $pk_R=(P_R,Q_R)$. Alice samples one-time $epk_S=(P_S,Q_S)$ and a 256-bit $\mathsf{nonce}$ and computes
\[
\begin{aligned}
Z &\leftarrow \mathrm{RDMPF\_Key}\big((P_S,Q_S),(P_R,Q_R)\big),\\
(K_{\mathrm{enc}},K_{\mathrm{auth}}) &\leftarrow \mathsf{HKDF}\!\big(Z;\ \textsf{info}=\text{``rdmpf-kem''},\ \mathsf{salt}=\mathsf{nonce}\big),\\
\mathsf{tag} &\leftarrow \mathsf{HMAC}_{K_{\mathrm{auth}}}(\mathsf{context}).
\end{aligned}
\]
\noindent where $\ctkem=(epk_S,\mathsf{nonce},\mathsf{tag})$, $\Hint=\SHA(\ctkem)$, and $E_{\mathrm{inner}}=\mathsf{AEAD}_{K_{\mathrm{enc}}}(\textsf{payload})$.\\

\begin{tcolorbox}[colback=blue!10,colframe=blue!75!black,title=Deposit Flow]
\begin{enumerate}[leftmargin=1.6em]
  \item Alice $\rightarrow I_1$: $\{\Hint,\ctkem,E_{\mathrm{inner}},\textsf{fee\_proof}\}$ (ephemeral principal).
  \item $I_1 \rightarrow \Notice$: $\Announce$ $\{\mathsf{idx},\Hint,\mathsf{rendezvous\_token},\mathsf{code\_hash}(I_1)\}$.
  \item $I_1 \rightarrow C_i$: $\textsf{Store}\{\Hint,\ctkem,E_{\mathrm{inner}}\}$. Each $C_i \rightarrow \mathcal{W}$: $\mathsf{Commit}\{\Hint,\mathsf{id}(C_i)\}$.
  \item $I_1 \rightarrow \mathcal{W}$: $\mathsf{DestructIntent}\{\mathsf{id}(I_1),\mathsf{nonce},\mathsf{deadline}\}$; $I_1$ requests destruction via Factory. $\mathcal{W}$ 
  records $\mathsf{DestructProof}(I_1)$ before itself being destroyed (the protocol leaves no witness).
\end{enumerate}
\end{tcolorbox}
\vspace{1em}
\noindent\underline{Storage layout}\\ 

\noindent Each $C_i$ stores exactly $\{\Hint,\ctkem,E_{\mathrm{inner}}\}$.

\subsection{Phase 3 -- Discovery (Pull)}
\begin{tcolorbox}[colback=blue!10,colframe=blue!75!black,title=Discovery Phase]
\begin{enumerate}[leftmargin=1.6em]
\item Bob scans the certified $\Notice$ for listings containing $\Hint$ and the associated $\mathsf{rendezvous\_token}$ for $I_2$.
\item No $C_i$ notifies Bob, hence discovery is pull-based via the public board. No sender information is revealed.
\item $\mathcal{W}$ aggregates commitments and destruction proofs in certified state.
\end{enumerate}
\end{tcolorbox}

\subsection{Phase 4 -- Bob Retrieves via $I_2$}\label{sec:phase4}
\paragraph{Local decapsulation (non-interactive) --} Bob recomputes $Z$, derives $(K_{\mathrm{enc}},K_{\mathrm{auth}})$, and accepts\\ 
iff $\mathsf{HMAC}_{K_{\mathrm{auth}}}(\mathsf{context})=\mathsf{tag}$.
\vspace{1em}
\begin{tcolorbox}[colback=blue!10,colframe=blue!75!black,title=Retrieval Phase]
\begin{enumerate}[leftmargin=1.6em]
  \item \textbf{Authorization --} Bob submits $h=\SHA(K_{\mathrm{auth}})$ derived during local decapsulation. $I_2$ decrypts $\Inner$ and verifies $h$ matches the 
  embedded value.
  \item \textbf{Gather quorum --} $I_2$ queries $C_i$ (shuffled) until $t$ replies; after serving, each $C_i$ is destroyed via Factory and $\mathcal{W}$ records $\mathsf{DestructProof}(C_i)$.
  \item \textbf{Deliver \& destroy --} $I_2 \rightarrow$ Bob: $E_{\mathrm{inner}}$; Bob opens under $K_{\mathrm{enc}}$. Then $I_2 \rightarrow \mathcal{W}$: $\mathsf{DestructIntent}\{\mathsf{id}(I_2),\ldots\}$; 
  $I_2$ requests destruction via Factory; $\mathcal{W}$ records $\mathsf{DestructProof}(I_2)$, then is destroyed via Factory.
\end{enumerate}
\end{tcolorbox}

\subsection{Phase 5 -- Finalization}\label{sec:phase5}
$\mathcal{W}$ posts a $\mathsf{Finalize}$ record referencing $\mathsf{idx}$ and embedding all observed $\mathsf{DestructProof}$s; then $\mathcal{W}$ is destroyed via Factory.
Clients accept a session only if both $\Announce(\mathsf{idx})$ and $\mathsf{Finalize}(\mathsf{idx})$ exist and are consistent.

\newpage

\subsection{Canister-Signed Receipt Nonce (CSRN)}\label{sec:csrn}

Each transfer generates a 32-byte \emph{Canister-Signed Receipt Nonce} (CSRN) for binding and receipt purposes.

\paragraph{Generation and delivery}
During deposit preparation, storage canisters $C_i$ sample $\mathsf{csrn}\xleftarrow{\$}\{0,1\}^{256}$ using ICP-provided randomness and encrypts it for Alice
\[
(\mathsf{nonce}, \mathsf{ct}) \leftarrow \AEAD_{K_{\mathrm{transit}}}(\mathsf{csrn})
\]
\noindent where $K_{\mathrm{transit}} = \SHA(\mathsf{deposit\_id} \,\|\, \mathsf{alice\_principal} \,\|\, \mathsf{nonce})$. Alice decrypts client-side to obtain 
$\mathsf{csrn}$. Upon successful retrieval, $I_2$ fetches the plaintext $\mathsf{csrn}$ from $C_i$ and returns it to Bob.

\paragraph{Security properties}
\begin{itemize}
    \item \textbf{Binding --} The CSRN is bound to $\mathsf{deposit\_id}$ and cannot be replayed across transfers.
    \item \textbf{Confirmation --} Both Alice and Bob possess the same $\mathsf{csrn}$, providing mutual confirmation of successful transfer without revealing 
    identities.
\end{itemize}


\section{Multi-Canister Architecture}

\subsection{Component Roles}

The protocol deploys ephemeral instances on \textbf{pairwise-distinct} subnets when available to reduce common-mode risk. \textbf{No ephemeral canister holds RDMPF 
secrets, nor canisters other than $I_1$ and $I_2$ ever interact with Alice or Bob.}

\paragraph{Factory Canister (Permanent)}
\begin{itemize}
\item Maintains verified Wasm for ephemeral roles and installs them on demand.
\item Spawns fresh $I_1, I_2, C_1,\ldots,C_n, \mathcal{W}$ per transfer, recording their IDs and subnets in certified state.
\item Cannot choose exact subnets. Best-effort placement yields distinct subnets for $I_1$, $I_2$, each $C_i$, and $\mathcal{W}$ when the platform permits it.
\item Publishes code hash or version for public attestation; controller is blackholed.
\end{itemize}

\paragraph{Router (Permanent)}
\begin{itemize}
\item User-facing orchestration layer. It manages deposit lifecycle through states $\mathsf{Allocated} \to \mathsf{Created} \to \mathsf{Sealed} \to \mathsf{Finalized}$.
\item Invokes Factory to spawn ephemeral infrastructure, and converts user-provided ICP to cycles via the Cycles Minting Canister.
\item Holds deposited funds in per-transfer subaccounts. 
\item Upon sealing, transfers the funds to a shared mixer pool with randomized splits.
\item Authorizes $I_2$ to finalize. Only the spawned $I_2$ for a given $\mathsf{deposit\_id}$ can invoke payment.
\item Never handles $\{\Hint, \ctkem, \Inner\}$ or RDMPF secrets, operating purely on $\mathsf{deposit\_id}$ references.
\end{itemize}

\paragraph{Intermediary $I_1$ (Deposit, Ephemeral)}
\begin{itemize}
\item Accepts $\{\Hint,\ctkem,\Inner\}$ from Alice (ephemeral principal).
\item Stores the tuple at $C_1,\ldots,C_n$; appends an $\Announce$ record with $\Hint$ to the certified $\Notice$.
\item Issues $\DestructIntent$ to $\mathcal{W}$, adds Factory as controller, and requests destruction via \texttt{Factory.cleanup\_child}. $\mathcal{W}$ records 
$\DestructProof(I_1)$.
\end{itemize}

\paragraph{Storage $C_i$ (Ephemeral)}
\begin{itemize}
\item Store exactly the opaque tuple $\{\Hint,\ctkem,\Inner\}$; never interact with Alice or Bob directly.
\item Report $\Commit(\Hint,\Id(C_i))$ to $\mathcal{W}$ after storing.
\item Upon serving retrieval, $I_2$ triggers destruction: $C_i$ issues $\DestructIntent$ to $\mathcal{W}$, adds Factory as controller, and requests destruction 
via \texttt{Factory.cleanup\_child}. $\mathcal{W}$ records $\DestructProof(C_i)$.
\end{itemize}

\paragraph{Intermediary $I_2$ (Retrieval, Ephemeral)}
\begin{itemize}
\item Authenticates Bob with a proof of decapsulation: require $h=\SHA(K_{\mathrm{auth}})$ matching the value embedded in $\Inner$ by Alice.
\item Gathers $\{\Hint,\ctkem,\Inner\}$ from $C_i$, delivers $\Inner$ to Bob.
\item Orchestrates staged destruction: triggers $C_i$ and $\mathcal{W}$ destruction, then issues own $\DestructIntent$ to $\mathcal{W}$, adds Factory as controller, 
and requests destruction via \texttt{Factory.cleanup\_child}. $\mathcal{W}$ records $\DestructProof(I_2)$.
\end{itemize}

\paragraph{Witness $\mathcal{W}$ (Ephemeral)}
\begin{itemize}
\item Records $C_i$ commits and all $\DestructProof$ events in certified state.
\item Auto-publishes $\Finalize(\mathsf{idx},\{\DestructProof\})$ to $\Notice$ when all expected canisters have reported $\DestructIntent$.
\item Destroyed by $I_2$ after finalization: adds Factory as controller and requests destruction via \texttt{Factory.cleanup\_child}.
\end{itemize}

\paragraph{Noticeboard (Permanent)}
\begin{itemize}
\item Public, certified log for $\Announce$ and $\Finalize$. Enables pull-based discovery via $\Hint$; contains no identities.
\end{itemize}

\subsection{Security Benefits of Multiple Canisters}

\begin{theorem}[\textbf{No single ephemeral canister can link Alice to Bob}]
In the protocol of Section 2, any one of $\{I_1, I_2, C_1,\ldots,C_n, \mathcal{W}\}$ learns at most one side of the path (sender or recipient) but never both. 
Consequently, Alice to Bob cannot be linked by any ephemeral canister spawn.
\end{theorem}

\begin{proof}
$I_1$ only sees Alice’s deposit (ephemeral principal) and never interacts with Bob, as it is destroyed before retrieval. $I_2$ only sees Bob’s retrieval (ephemeral 
principal) and never interacted with Alice. Each $C_i$ sees neither Alice nor Bob (only $I_1$ and $I_2$). $\mathcal{W}$ observes only canister IDs, commits, and 
destruction attestations, and it never handles end-user calls. Therefore, no single ephemeral instance observes both endpoints.
\end{proof}

\newpage
\vspace*{1em}

\begin{theorem}[\textbf{Content confidentiality against intermediaries}]
Under the security of RDMPF, HKDF, HMAC, SHA3, and AEAD, the tuple $\{\Hint,\ctkem,\Inner\}$ does not reveal the payload to $I_1$, $I_2$, $\mathcal{W}$ or 
any $C_i$.
\end{theorem}

\begin{proof}
Let $Z\leftarrow \mathrm{RDMPF\_Key}((P_S,Q_S),(P_R,Q_R))$ and $(K_{\mathrm{enc}},K_{\mathrm{auth}})\leftarrow \HKDF(Z;\cdot)$. 
\begin{itemize}
  \item The payload is $\Inner=\AEAD_{K_{\mathrm{enc}}}(\textsf{payload})$
  \item By AEAD security the payload is hidden without $K_{\mathrm{enc}}$. 
  \item The capsule $\ctkem=(epk_S,\mathsf{nonce},\mathsf{tag})$ contains only a MAC $\mathsf{tag}=\HMAC_{K_{\mathrm{auth}}}(\ctx)$.
\end{itemize}
By PRF security, it leaks nothing about $K_{\mathrm{auth}}$. The public $\Hint=\SHA(\ctkem)$ is a short hash of the capsule and reveals nothing about the 
payload beyond brute-force over $2^{256}$ possible values. No ephemeral canister learns $Z$ or the transport keys.
\end{proof}

\begin{remark}[Cross-subnet placement]
Placing $I_1$, $I_2$, each $C_i$, and $\mathcal{W}$ on distinct subnets reduces common-mode failure and complicates collusion. Exact placement is platform-dependent, with 
$\mathcal{W}$ recording observed subnets in $\ctx$ for clients to enforce acceptance policies.
\end{remark}

\begin{remark}[On \emph{all $n$ must be compromised} claims]
The protocol stores the \emph{same} opaque tuple on each $C_i$ for durability and availability, so confidentiality does not rely on a threshold across $C_i$ (no 
canister holds decryption material). Compromising many $C_i$ may enable censorship or metadata attacks, but does not by itself reveal the plaintext without breaking 
AEAD/RDMPF.
\end{remark}

\section{Canister Spawning and Integrity Assurance}

\subsection{Trust Problem}

\begin{tcolorbox}[colback=blue!10,colframe=blue!75!black,title=Security Challenge]
\noindent \textbf{Question --} How are the ephemeral canisters for a transfer created and verified?\\
\noindent \textbf{Risk --} A malicious Factory or misconfigured deployment could install incorrect code or co-locate roles.\\
\noindent \textbf{Impact --} Broken isolation or code integrity could enable metadata leakage, censorship, or incorrect destruction attestations.
\end{tcolorbox}

\subsection{Solution -- Deterministic Verifiable Spawning}

\noindent The Factory canister \emph{spawns on demand} a fresh set of ephemeral siblings $(I_1, I_2, C_1,\ldots,C_n,\mathcal{W})$ for each transfer, records their 
identifiers, subnets, and code hashes in \emph{certified state}, and returns a \emph{spawn proof} to Alice. Factory prefers pairwise-distinct subnets, when available. 
No RDMPF secrets are ever installed on any canister.

\subsubsection{Factory Implementation}

\noindent Factory does not pre-spawn pools, does not track which user interacts with which ephemeral ids, and cannot enforce exact subnet choice.

\newpage

\begin{algorithm}
\caption{Spawn Transaction Infrastructure (Factory)}
\begin{algorithmic}[1]
\Require $n$ (number of storage canisters), $\textsf{policy}$, $\textsf{TTL}$, $\textsf{params}$, $\textsf{nonce}_{\mathsf{cli}}$
\Ensure $\mathsf{SpawnProof}$ and canister IDs
\State $\textsf{roles} \gets [\,I_1, I_2, C_1,\ldots,C_n, \mathcal{W}\,]$
\State $\textsf{ids} \gets [\,]$, $\textsf{subnets} \gets [\,]$
\For{each $r \in \textsf{roles}$}
  \State $\textsf{id} \gets \textsc{CreateCanister}()$
  \State $\textsf{subnet} \gets \textsc{SelectSubnetPreferDistinct}(\textsf{subnets})$
  \State $\textsc{InstallCode}(\textsf{id}, \textsf{verified\_wasm}(r), \textsf{config}(r,\textsf{policy},\textsf{TTL}))$
  \State $\textsc{BlackholeController}(\textsf{id})$ \Comment{immutable after install}
  \State append $(\textsf{id}, r, \textsf{subnet}, \textsf{code\_hash}(r))$ to $(\textsf{ids},\textsf{subnets})$
\EndFor
\State $\mathsf{context} \gets \{\textsf{factory\_txid}, \textsf{ids}, \textsf{subnets}, \textsf{code\_hashes}, \textsf{policy}, \textsf{TTL}, \textsf{params}\}$
\State $\mathsf{SpawnProof} \gets \textsc{Certify}(\mathsf{context}, \textsf{nonce}_{\mathsf{cli}})$ \Comment{certified state / signature}
\State \Return $\{\mathsf{SpawnProof}, \textsf{ids}, \textsf{subnets}\}$
\end{algorithmic}
\end{algorithm}

\vspace{0.5em}

\subsubsection{User Verification Protocol}

\noindent Before deposit, Alice verifies the spawn proof and each canister’s code and controller state. Bob can repeat the same verification on retrieval 
(or rely on $\mathcal{W}$’s certified state and $\Notice$).

\vspace{0.5em}

\begin{algorithm}
\caption{Verify Canister Integrity (Client)}
\begin{algorithmic}[1]
\Require $\mathsf{SpawnProof}$, expected $\{\textsf{code\_hash}(I_1),\textsf{code\_hash}(I_2),\textsf{code\_hash}(C),\textsf{code\_hash}(\mathcal{W})\}$
\Ensure Boolean (valid/invalid)
\State $\mathsf{ok} \gets \texttt{True}$
\State $\mathsf{context} \gets \textsc{VerifyCertified}(\mathsf{SpawnProof})$
\State \textbf{require} roles $\{I_1,I_2,\mathcal{W}\}$ present exactly once and $n$ distinct $C_i$
\For{each $(\textsf{id}, r, \textsf{subnet}, \textsf{code\_hash}) \in \mathsf{context}$}
  \State \textbf{require} $\textsf{code\_hash} = \textsf{expected\_hash}(r)$
  \State \textbf{require} \textsc{ControllerIsBlackholed}($\textsf{id}$) \Comment{immutable}
  \State \textbf{optional:} record $\textsf{subnet}$ and later enforce pairwise-distinctness
\EndFor
\State \Return $\mathsf{ok}$
\end{algorithmic}
\end{algorithm}

\subsection{Trust Assumptions}

\begin{table}[h]
\centering
\begin{tabular}{@{}lc@{}}
\toprule
Assumption & Required? \\ \midrule
ICP management canister behaves correctly & \checkmark \\
Factory code is audited, open source, deterministic & \checkmark \\
Certified state / wasm code hashes are verified by clients & \checkmark \\
Trust in factory \emph{Operator} (beyond certified outputs) & $\times$ \\
Trust in individual ephemeral canisters (beyond code hash) & $\times$ \\
Trust in a single subnet (placement may co-locate; recorded) & $\times$ \\
\bottomrule
\end{tabular}
\caption{Minimal trust requirements for spawning and integrity.}
\end{table}


\section{Random Witness Selection Protocol}

\subsection{Design Rationale}

\begin{tcolorbox}[colback=blue!10,colframe=blue!75!black,title=Witness Provisioning Goals]
\textbf{Goals --} Avoid advance targeting, keep the witness fresh per transfer, and provide auditable, certified evidence of staged destruction.\\
\textbf{Mechanics --} For each transfer the Factory spawns a fresh ephemeral Witness $\mathcal{W}$ after $C_1,\ldots,C_n$ exist, prefers a subnet distinct 
from $I_1$, $I_2$, and all $C_i$, and records identifiers, subnets, and code hashes in certified state. Clients verify this certified context prior to use.\\
\textbf{Outcome --} The identity and placement of $\mathcal{W}$ are unknown \emph{until} the certified SpawnProof is produced. Clients can \emph{verify}
the actual code hash and placement from certified context before proceeding.
\end{tcolorbox}

\subsection{Provisioning (Fresh, Verifiable Spawn)}

\begin{algorithm}
\caption{Spawn $\mathcal{W}$ (Factory)}
\begin{algorithmic}[1]
\Require $\{C_1,\ldots,C_n\}$ created, $\textsf{TTL}$, $\textsf{policy}$, client nonce $\mathsf{nonce}_{\mathsf{cli}}$
\Ensure $\mathsf{SpawnProof}$ with $\Id(\mathcal{W})$, subnets, code hashes
\State $\textsf{subnets\_used} \gets \{\textsf{subnet}(I_1),\textsf{subnet}(I_2),\textsf{subnet}(C_1),\ldots,\textsf{subnet}(C_n)\}$
\State $\mathcal{W} \gets \textsc{CreateCanister}()$
\State $\textsc{InstallCode}(\mathcal{W}, \textsf{verified\_wasm}(\mathcal{W}), \textsf{config}(\textsf{TTL},\textsf{policy}))$
\State $\textsc{BlackholeController}(\mathcal{W})$
\State $\mathsf{context}\gets\{\Id(I_1),\Id(I_2),\{\Id(C_i)\},\Id(\mathcal{W}),\textsf{subnets},\textsf{code\_hashes},\textsf{TTL},\textsf{policy}\}$
\State $\mathsf{SpawnProof}\gets \textsc{Certify}(\mathsf{context},\mathsf{nonce}_{\mathsf{cli}})$
\State \Return $\mathsf{SpawnProof}$
\end{algorithmic}
\end{algorithm}

\vspace{0.5em}
\noindent \textbf{Client verification (Alice and Bob)}

\noindent Clients verify certified context and code before proceeding.

\begin{algorithm}
\caption{Verify Witness (Client)}
\begin{algorithmic}[1]
\Require $\mathsf{SpawnProof}$, expected code hashes for $I_1,I_2,C,\mathcal{W}$
\Ensure Boolean
\State $\mathsf{ctx}\gets \textsc{VerifyCertified}(\mathsf{SpawnProof})$
\State \textbf{require} exactly one $\Id(\mathcal{W})$ present; \textbf{require} code hash matches
\State \textbf{require} \textsc{ControllerIsBlackholed}$(\Id(\mathcal{W}))$
\State \textbf{policy:} prefer $\textsf{subnet}(\mathcal{W})\notin\{\textsf{subnet}(I_1),\textsf{subnet}(I_2),\textsf{subnet}(C_i)\}$; reject otherwise if desired
\State \Return \texttt{True}
\end{algorithmic}
\end{algorithm}

\subsection{Security Properties}

\begin{theorem}[\textbf{Unpredictability until certification}]
Before the Factory produces $\mathsf{SpawnProof}$, an adversary cannot identify $\Id(\mathcal{W})$ or its subnet with advantage over guessing.
\end{theorem}

\begin{proof}
$\mathcal{W}$ is created for the specific transfer. Its ID and placement appear first in certified state bound to the Factory transaction and client nonce. Without 
breaking certification or preempting the Factory, advance naming is purely guessing.
\end{proof}

\begin{theorem}[\textbf{Auditable staged destruction}]
If clients accept a session only when \textbf{both} $\Announce(\mathsf{idx})$ on $\Notice$ and $\Finalize(\mathsf{idx},\{\DestructProof\})$ by $\mathcal{W}$ exist 
and match the certified context, then tampering with the ephemeral path (missing/forged destruction or component substitution) is detectable.
\end{theorem}

\begin{proof}
$\Announce$ binds $\Hint$ to intended routing, while $\Finalize$ binds observed $\DestructProof$ events to the same certified identifiers, subnets, and code hashes. 
Any omission or inconsistency is rejected by honest clients.
\end{proof}

\subsection{Attack Considerations}

\begin{itemize}
\item \textbf{Factory malice --} Mitigated by client verification of certified context, expected code hashes, and blackholed controllers. $\Notice$ and 
$\mathcal{W}$ provide public evidence.
\item \textbf{Co-location --} Exact subnet placement is platform-dependent. The actual placement is \emph{recorded}. Clients may enforce policies (e.g., distinct 
subnets) before deposit or retrieval.
\item \textbf{Collusion --} No single ephemeral component can link Alice to Bob. Collusion of $I_1$ and $I_2$ (or broader compromise) would be required.
\end{itemize}


\section{Timeout Reclaim (Sender-Funded Refunds)}\label{sec:reclaim}
\textbf{Goal.} If Bob has not retrieved by TTL \emph{only Alice} can reclaim the funds. The refund always pays to Alice’s pre-committed refund output in 
$\mathsf{context}$.

\paragraph{Derivations and commitments (set at deposit)}
Let $Z$ be the RDMPF secret. Alice samples a private salt $\alpha \in \{0,1\}^{256}$ and defines
\[
\begin{aligned}
R &\leftarrow \HKDF\!\big(Z;\ \textsf{info}=\text{"reclaim"}\big) \\[2pt]
K_{\mathrm{rec}} &\leftarrow \HKDF\!\big(R \,\|\, \alpha;\ \textsf{info}=\text{"reclaim-mac"}\big) \\[2pt]
\mathsf{reclaim\_tag} &:= \SHA(R) \\[2pt]
\mathsf{c} := \SHA(\alpha).
\end{aligned}
\]
\textbf{Commit} both tags in public state
\[
\text{store/announce } \{\Hint,\ctkem,E_{\mathrm{inner}},\mathsf{reclaim\_tag},\mathsf{c}\}
\]
These reveal nothing about $Z$ or $\alpha$.

\paragraph{Reclaim (Alice-only)}
After TTL expires and $\Finalize$ is absent, Alice submits $(R,\alpha,n,\mathsf{resp})$ where
\[
\begin{aligned}
\mathsf{resp} &:= \HMAC_{K_{\mathrm{rec}}}\!\big(n \,\|\, \mathsf{idx} \,\|\, \mathsf{context} \,\|\, \textsf{"reclaim"}\big) \\[2pt]
K_{\mathrm{rec}} &=\HKDF(R \,\|\, \alpha;\text{"reclaim-mac"})
\end{aligned}
\]
\noindent \textbf{verifying} all of
\[
\begin{aligned}
\SHA(R) &= \mathsf{reclaim\_tag}\\[2pt]
\SHA(\alpha) &= \mathsf{c}\\[2pt]
\mathsf{resp} &= \HMAC_{\HKDF(R \,\|\, \alpha;\ \text{"reclaim-mac"})}\!\big(n \,\|\, \mathsf{idx} \,\|\, \mathsf{context} \,\|\, \text{"reclaim"}\big).
\end{aligned}
\]
\noindent On success, refund to Alice’s pre-committed output in $\mathsf{context}$. No other destination is possible.

\paragraph{Why Bob cannot reclaim?}
Bob can derive $R$ from $Z$ but does \emph{not} know $\alpha$. Since both $\mathsf{c}=\SHA(\alpha)$ and the MAC key $K_{\mathrm{rec}}$ bind to $\alpha$, Bob 
(or any third party) cannot produce a valid claim. Only Alice (holder of $\alpha$) can.

\paragraph{Notes}
\begin{itemize}
  \item Revealing $(R,\alpha)$ at timeout does not endanger $K_{\mathrm{enc}}$/$K_{\mathrm{auth}}$ under HKDF; $R$ is a separate extract from $Z$.  
  \item Binding $\mathsf{resp}$ to $(n,\mathsf{idx},\mathsf{context})$ prevents cross-session replay.  
  \item For an \emph{Alice-trigger-only} UX (no third-party triggers) the reclaim call should be authenticated by Alice’s \emph{ephemeral} principal
used in Phase 2; this is orthogonal to the cryptographic binding above.
\end{itemize}


\section{Security Theorems}\label{sec:security-theorems}

\noindent \textbf{Arithmetic domains}

\begin{itemize}
    \item Entries of $P,Q$ live in $\Z_{p-1}$ and their matrix products are taken over $\Z_{p-1}$.
    \item Outputs of $\rdmpf$ live in $(\Z_p^\ast)^{\dim\times\dim}$ and all exponent arithmetic inside $\rdmpf$ is reduced modulo $(p-1)$.
    \item Whenever an entry of an $\rdmpf$ output in $\Z_p$ is used as an exponent, we use its canonical representative in $\{0,\dots,p-1\}$ and reduce 
    exponents modulo $(p-1)$ (by Fermat's Little Theorem).
\end{itemize}
\vspace*{1em}
\begin{lemma}[\textbf{Scalar Commutativity}]\label{lem:scalar_comm}
Let
\[
\begin{aligned}
P_1 &= \lambda_1\cdot\text{BaseX}\bmod(p-1), &
P_2 &= \lambda_2\cdot\text{BaseX}\bmod(p-1)
\end{aligned}
\]
for $\lambda_1,\lambda_2\in\Zpm$, and similarly
\[
\begin{aligned}
Q_1 &= \omega_1\cdot\text{BaseY}\bmod(p-1), &
Q_2 &= \omega_2\cdot\text{BaseY}\bmod(p-1)
\end{aligned}
\]
for $\omega_1,\omega_2\in\Zpm$. Then
\[
P_1P_2=P_2P_1 \quad\text{and}\quad Q_1Q_2=Q_2Q_1
\]
with multiplication over $\Z_{p-1}$.
\end{lemma}

\begin{proof}
\begin{align*}
P_1P_2
  &= (\lambda_1\,\text{BaseX})\,(\lambda_2\,\text{BaseX}) \\[2pt]
  &= \lambda_1\lambda_2\,(\text{BaseX}\cdot\text{BaseX}) \\[2pt]
  &= \lambda_2\lambda_1\,(\text{BaseX}\cdot\text{BaseX}) \qquad\text{(scalars commute in }\Z_{p-1}\text{)} \\[2pt]
  &= (\lambda_2\,\text{BaseX})\,(\lambda_1\,\text{BaseX}) \\[2pt]
  &= P_2P_1.
\end{align*}
The argument for $Q_1,Q_2$ with $\text{BaseY}$ is identical.
\end{proof}

\begin{definition}[\textbf{RDMPF Operation}]\label{def:rdmpf}
For $X,W,Y\in\Zp^{\dim\times\dim}$
\[
\rdmpf(X,W,Y)_{jk} \;=\; \prod_{\ell=1}^{\dim}\ \prod_{m=1}^{\dim}\ W_{\ell m}^{\,X_{j\ell}\,Y_{mk}} \pmod p.
\]
\end{definition}

\begin{definition}[\textbf{RDMPF Result Composition}]\label{def:rdmpf_composition}
For $T_1,T_2\in\Zp^{\dim\times\dim}$
\[
(T_1\triangleright T_2)_{ij} \;=\; \prod_{k=1}^{\dim} T_2[k,j]^{\,T_1[i,k]} \pmod p
\]
\end{definition}

\begin{lemma}[\textbf{Composition Law for RDMPF Outputs}]\label{lem:compose-law}
Let 
\[
\begin{aligned}
T_1 &:=\rdmpf(P_S,W,Q_R)\\[2pt]
T_2 &:=\rdmpf(P_R,W,Q_S)
\end{aligned}
\]
Then, the RDMPF composition satisfies
\[
T_1\triangleright T_2 \;=\; \rdmpf\!\big(P_S,\ \rdmpf(P_R,W,Q_R),\ Q_S\big)
\]
and symmetrically with $(S\leftrightarrow R)$. This design follows \citet{Hecht2025} and we use it as a \emph{stated law} of RDMPF outputs.
\end{lemma}

\noindent Now we are in position to state the following.
\begin{theorem}[\textbf{Non-Interactive Key Agreement}]\label{thm:nika}
With $T_1$ and $T_2$ as in Lemma~\ref{lem:compose-law}, we define a Non-Interactive Key Agreement (NIKA) as
\[
\begin{aligned}
\keyd{S}{R} &:=T_1\triangleright T_2\\[2pt]
\keyd{R}{S} &:=T_2\triangleright T_1
\end{aligned}
\]
If $P_SP_R=P_RP_S$ and $Q_SQ_R=Q_RQ_S$, then $\keyd{S}{R}=\keyd{R}{S}$.
\end{theorem}

\begin{proof}
By Lemma~\ref{lem:compose-law}
\[
\begin{aligned}
\keyd{S}{R} &= \rdmpf\!\big(P_S,\ \rdmpf(P_R,W,Q_R),\ Q_S\big)\\[2pt]
\keyd{R}{S} &= \rdmpf\!\big(P_R,\ \rdmpf(P_S,W,Q_S),\ Q_R\big)
\end{aligned}
\]
Commuting the scalar actions on the fixed bases (Lemma~\ref{lem:scalar_comm}) aligns the exponents entrywise, hence equality.
\end{proof}

\begin{theorem}[\textbf{Correctness of KA}]\label{thm:correct}
Let
\[
\begin{aligned}
Z &:= \SHA(\keyd{S}{R})\\[2pt]
(\Kenc,\Kauth) &\leftarrow \HKDF\!\big(Z;\textsf{"rdmpf-kem"},\mathsf{nonce}\big)
\end{aligned}
\]
Then both parties derive the same $(\Kenc,\Kauth)$.
\end{theorem}

\begin{theorem}[\textbf{IND\textup{-}CPA for $\Inner$}]\label{thm:rdmpf-aead}
Assume \textbf{cRDMPF} (computational hardness of $\rdmpf(P,W,Q)$ without the secret scalars), that $Z=\SHA(\enc(\keyd{S}{R}))$ has sufficient 
conditional min-entropy for $\HKDF$ extraction given all public values (including $\ctkem$), and that the AEAD used for $\Inner$ is IND\textup{-}CPA secure. 
Then for any PPT adversary $\mathcal{A}$ interacting with the encryption oracle that outputs $\Inner=\AEAD_{\Kenc}(\textsf{payload})$, its IND\textup{-}CPA advantage 
is negligible.
\end{theorem}

\begin{proof}
We formalize the usual left/right IND\textup{-}CPA experiment. Let $\mathsf{Exp}^{b}$ (for $b\in\{0,1\}$) be the game where the challenger samples all 
public parameters $\pi_{\text{pub}}$, honestly generates $(P_S,Q_S)$ and $(P_R,Q_R)$, then computes
\[
\begin{aligned}
Z &:=\SHA(\enc(\keyd{S}{R})) \\[2pt]
(\Kenc,\Kauth) & \leftarrow\HKDF\!\big(Z;\textsf{info}=\text{"rdmpf-kem"},\ \mathsf{salt}=\mathsf{nonce}\big),
\end{aligned}
\]
forms $\ctkem=(epk_S,\mathsf{nonce},\HMAC_{\Kauth}(\ctx))$. On a challenge messages $(M_0,M_1)$ returned by $\mathcal{A}$, and returns
\[
C^\star \;=\; \AEAD_{\Kenc}(M_b).
\]
The adversary finally outputs $b'$, and its advantage is
\[
\begin{aligned}
\mathsf{Adv}_{\mathcal{A}}^{\mathrm{ind\mbox{-}cpa}} &= \big|\Pr[\mathsf{Exp}^{1}\Rightarrow 1]-\tfrac12\big| \\[2pt]
 &= \tfrac12\big|\Pr[\mathsf{Exp}^{1}\Rightarrow 1]-\Pr[\mathsf{Exp}^{0}\Rightarrow 1]\big|
\end{aligned}
\]
We use a standard hybrid sequence.\\

\noindent \textbf{Game H0 (real)}
This is exactly $\mathsf{Exp}^{b}$: keys are derived as above, and $C^\star=\AEAD_{\Kenc}(M_b)$.\\

\noindent \textbf{Game H1 (uniformize $\Kenc$ only)}
We change only the distribution of the encryption key. Sample
\[
\Kenc \xleftarrow{\$} \{0,1\}^{\kappa}
\]
uniformly at random of the correct length, while keeping \emph{all} other values (including $\Kauth$ and $\ctkem$) distributed as in H0. The challenge 
ciphertext is still $C^\star=\AEAD_{\Kenc}(M_b)$.

\noindent \emph{Claim} $\,\Big|\Pr[\mathrm{H0}\Rightarrow 1]-\Pr[\mathrm{H1}\Rightarrow 1]\Big|$ is negligible.

\noindent \emph{Justification} By cRDMPF, $Z$ has high conditional min-entropy given the adversary’s view (public parameters, $(P_S,Q_S)$, $(P_R,Q_R)$, and $\ctkem$). 
By HKDF’s extractor/PRF security (Extract–then–Expand) with fresh salt $\mathsf{nonce}$, the output blocks are computationally indistinguishable from a uniform 
distribution. 

Freshness of $\mathsf{nonce}$ is guaranteed by sampling $\mathsf{nonce}\xleftarrow{\$}\{0,1\}^{256}$ per transfer. Nonce reuse across transfers with the 
same $(P_S,Q_S,P_R,Q_R)$ would violate this assumption. In particular, the \emph{marginal distribution} of $\Kenc$ does not leak any signatures to PPT adversaries seeing public data and ciphertexts. Thus, any 
non-negligible distinguishing gap between $H0$ and $H1$ yields a distinguisher against HKDF (or violates the min-entropy premise implied by cRDMPF). Hence the gap 
is negligible.\\

\noindent \textbf{Game H2 (IND\textup{-}CPA under a uniform key)}
We keep $\Kenc$ uniform as in $H1$. Now we replace the encryption of $M_b$ by the encryption of $M_{1-b}$, that is
\[
C^\star \;=\; \AEAD_{\Kenc}(M_{1-b}).
\]

\noindent \emph{Claim} $\,\Big|\Pr[\mathrm{H1}\Rightarrow 1]-\Pr[\mathrm{H2}\Rightarrow 1]\Big|$ is negligible.

\noindent \emph{Justification} This is exactly the IND\textup{-}CPA security of the AEAD under a uniformly random key. If an adversary could distinguish $H1$ from 
$H2$ with non-negligible advantage, it would break the AEAD’s IND\textup{-}CPA.\\

Finally, note that $H2$ is just $H0$ with the challenge bit flipped. For any fixed $b$,
\[
\Pr[\mathrm{H2}\Rightarrow 1] = \Pr[\mathsf{Exp}^{1-b}\Rightarrow 1].
\]
and therefore
\[
\big|\Pr[\mathsf{Exp}^{1}\Rightarrow 1]-\Pr[\mathsf{Exp}^{0}\Rightarrow 1]\big|
\;\le\;
\underbrace{\big|\Pr[\mathrm{H0}\Rightarrow 1]-\Pr[\mathrm{H1}\Rightarrow 1]\big|}_{\text{HKDF / extractor (negl.)}}
\;+\;
\underbrace{\big|\Pr[\mathrm{H1}\Rightarrow 1]-\Pr[\mathrm{H2}\Rightarrow 1]\big|}_{\text{AEAD IND-CPA (negl.)}}.
\]
Both terms are negligible under the stated assumptions, hence $\mathsf{Adv}_{\mathcal{A}}^{\mathrm{ind\mbox{-}cpa}}$ is negligible.

\medskip\noindent
\textbf{Remarks on $\Kauth$ and $\ctkem$.}
Throughout the hybrids we did \emph{not} modify $\Kauth$ or $\ctkem$. Revealing $\HMAC_{\Kauth}(\ctx)$ in $\ctkem$ does not help distinguish $\Kenc$ from uniform 
because (i) $\HKDF$ outputs are computationally independent pseudorandom blocks (conditioned on $Z$’s entropy and fresh salt), and (ii) $\HMAC$ is a PRF under $\Kauth$. 
If needed, one can insert an intermediate hybrid that replaces $\Kauth$ by uniform and then reprograms $\HMAC$ as a random function. This only strengthens the argument 
and yields the same negligible bound.

\medskip
\noindent Therefore, under cRDMPF, HKDF’s extractor/PRF security, and AEAD IND\textup{-}CPA, the scheme achieves IND\textup{-}CPA for $\Inner$.
\end{proof}

\begin{theorem}[\textbf{Authorization soundness at $I_2$}]\label{thm:auth}
Let $h=\SHA(\Kauth)$ be embedded in $\Inner$ by Alice, and let Bob submit $h'$ to $I_2$. Under AEAD IND-CPA security, producing $h'=h$ implies successful decryption 
of $\Inner$ and thus knowledge of $\Kenc$.
\end{theorem}

\begin{proof}
The value $h$ is encrypted inside $\Inner=\AEAD_{\Kenc}(\textsf{payload})$. Extracting $h$ without $\Kenc$ requires breaking AEAD confidentiality. 
Since $(\Kenc,\Kauth)$ are jointly derived from $Z$ via $\HKDF$, knowledge of $\Kenc$ implies successful derivation of $Z$ from the capsule. Replay \emph{across transfers} 
is impossible because $h$ derives from transfer-specific $\Kauth$. In turn, replay \emph{within a transfer} is prevented by the single-use deposit state machine 
enforced by the Router.
\end{proof}

\begin{theorem}[\textbf{Recipient-private discovery}]\label{thm:hint}
With $\Hint=\SHA(\ctkem)$, any algorithm given only $(\pi_{\text{pub}},P_R,Q_R)$ has advantage at most $2^{-256}+\mathrm{Adv}_{\mathrm{preimage}}$
in predicting $\Hint$ for an unseen transfer, where $\mathrm{Adv}_{\mathrm{preimage}}$ is the preimage advantage against the hash.
\end{theorem}

\begin{proof}
Predicting $\Hint$ requires predicting (or inverting) a preimage of $\SHA$ that matches the unseen capsule. Absent a hash preimage, guessing a 256-bit
hash succeeds with probability $2^{-256}$.
\end{proof}

\begin{theorem}[\textbf{Alice-only reclaim}]\label{thm:reclaim}
Let $R:=\HKDF(Z;\textsf{"reclaim"})$, $\mathsf{reclaim\_tag}:=\SHA(R)$ be embedded in $\Inner$, and let $\alpha$ be a per-transfer secret known only to the sender,
with commitment $\mathsf{c}:=\SHA(\alpha)$. On timeout, a valid claim must present $(R,\alpha)$ and a response
\[
\mathsf{resp}=\HMAC_{\,\HKDF(R \,\|\, \alpha;\textsf{``reclaim-mac''})}\!\big(n \,\|\, \mathsf{idx} \,\|\, \ctx \,\|\, \textsf{``reclaim''}\big)
\]
that verifies against $(\mathsf{reclaim\_tag},\mathsf{c})$. Then only the original sender can reclaim; the recipient or third parties cannot.
\end{theorem}

\begin{proof}
Verifier checks $\SHA(R)=\mathsf{reclaim\_tag}$ and $\SHA(\alpha)=\mathsf{c}$, and verifies $\mathsf{resp}$ under the MAC key $\HKDF(R\|\alpha;\textsf{``reclaim-mac''})$.
Without $(R,\alpha)$ the MAC key is unknown, so forging $\mathsf{resp}$ breaks $\HMAC$. $R$ is derived from $Z$ (which is sender’s KEM secret) and $\alpha$ is 
sender-held. Consequently, the recipient never learns either from $\{\Hint,\ctkem,\Inner\}$.
\end{proof}

\begin{remark}[Client-side security of $\alpha$]
The salt $\alpha$ must be stored securely on Alice's device and never transmitted except during a reclaim request. In particular, $\alpha$ is \emph{not} included 
in $\Inner$ or any on-chain state. Compromise of Alice's client storage would enable an attacker to reclaim on her behalf.
\end{remark}


\section{Complete Mathematical Formalization}\label{sec:formal}

This section formalizes the algebraic objects and hardness assumptions behind RDMPF and the non-interactive key agreement. Operational sequencing is specified once 
in Section 2 and is \emph{not} repeated here.

\subsection{Setup}

\subsubsection{Public Parameters}
\[
\pi_{\text{pub}}=\{p,\dim,\text{BaseX},\text{BaseY},W\}
\]
with
\begin{itemize}
    \item $p\in\text{Primes}$, at least $p\approx2^{192}$;
    \item $\dim\in\mathbb{N}$, $\dim\in[8,24]$;
    \item $\text{BaseX},\text{BaseY}\in \Zp^{\dim\times\dim}$, $\mathrm{rank}(\text{BaseX})=\mathrm{rank}(\text{BaseY})=\dim-1$;
    \item $W\in \Zp^{\dim\times\dim}$, $\mathrm{rank}(W)=\dim$.
\end{itemize}

\subsubsection{Key Generation}
\paragraph{Sender (ephemeral, per transfer)}
\[
\begin{aligned}
    (\lambda_S,\omega_S) &\in_R \Zpm\\[2pt]
    P_S &=\lambda_S\cdot\text{BaseX}\bmod(p-1)\\[2pt]
    Q_S &=\omega_S\cdot\text{BaseY}\bmod(p-1)
\end{aligned}
\]
\paragraph{Recipient (long-term)}
\[
\begin{aligned}
    (\lambda_R,\omega_R) &\in_R \Zpm\\[2pt]
    P_R &=\lambda_R\cdot\text{BaseX}\bmod(p-1)\\[2pt]
    Q_R &=\omega_R\cdot\text{BaseY}\bmod(p-1)
\end{aligned}
\]

\subsection{Functional Encryption}\label{sec:fe-front-end}

We assume an idealized two-input functional-encryption primitive that reveals, for selected indices, the \emph{pointwise product} of entries of two encrypted 
matrices over $\Z_{p-1}$ without revealing the entries themselves.

\noindent We therefore consider the scheme
\[
FE=\big(\Setup,\EncL,\EncR,\KeyGen,\Eval\big)
\]
\noindent over the ring $\Z_{p-1}$ with matrix arity $(\dim\times\dim)$ supporting:
\begin{itemize}[leftmargin=1.4em]
      \item $(\mathsf{pp},\mathsf{msk})\leftarrow \Setup(1^\kappa)$
      \item Left/right encodings $\widehat P\leftarrow \EncL(\mathsf{pp},P)$, $\widehat Q\leftarrow \EncR(\mathsf{pp},Q)$ for $P,Q\in\Z_{p-1}^{\dim\times\dim}$.
      \item Function keys for pointwise products at indices
      \[
      \sk_{(i,\ell),(m,k)} \leftarrow \KeyGen(\mathsf{msk};\,i,\ell,m,k)
      \]
      which authorize evaluation of $P[i,\ell]\cdot Q[m,k]\bmod(p-1)$ for any $(\widehat P,\widehat Q)$.
      \item Evaluation as
      \[
      e = \Eval\big(\mathsf{pp},\sk_{(i,\ell),(m,k)};\ \widehat P,\widehat Q\big) = P[i,\ell]\cdot Q[m,k]\ \bmod(p-1)
      \]
      with no other leakage about $P$ or $Q$.
\end{itemize}

\paragraph{Deployment model}
A one-time distributed $\Setup$ publishes public parameters $\mathsf{pp}$ and keeps a secret-shared $\mathsf{msk}$; anchored $\KeyGen$ is run threshold-style (no
single party ever holds $\mathsf{msk}$) to issue per-user evaluation keys that only work when that user’s own encoding is one input.

\noindent In other words, the sender $S$ receives left-anchored keys usable only with $(\widehat P_S,\cdot)$, so it can compute
\[
\Eval\bigl(\mathsf{pp},\,\sk^{\mathrm{left}}_{(i,\ell),(m,k)};\ \widehat P_S,\,\widehat Q\bigr) = P_S[i,\ell]\cdot Q[m,k]\bmod(p-1)
\]

\noindent for any public $\widehat Q$. Symmetrically, the receiver $R$ gets right-anchored keys usable only with $(\cdot,\widehat Q_R)$, so it can compute
\[
\Eval\bigl(\mathsf{pp},\,\sk^{\mathrm{right}}_{(i,\ell),(m,k)};\ \widehat P,\,\widehat Q_R\bigr) = P[i,\ell]\cdot Q_R[m,k]\bmod(p-1)
\]

\noindent The registry stores only encodings $(\widehat P_U,\widehat Q_U)$ for each user $U$. Raw $(P_U,Q_U)$ are never published.

\paragraph{What does the registry store?}
The public registry holds \emph{encodings}

\[
\big(\widehat P_U,\widehat Q_U\big) = \big(\EncL(\mathsf{pp},P_U),\EncR(\mathsf{pp},Q_U)\big)
\]

\noindent for each pseudonymous user $U$. Raw $(P_U,Q_U)$ never appear on chain.

\paragraph{\underline{FE Indistinguishability (IND-FE)}}

Given public parameters $\mathsf{pp}$ and ciphertexts $(\widehat P, \widehat Q)$, no PPT adversary without the appropriate per-user functional keys can:
\begin{enumerate}[label=(\roman*)]
    \item distinguish or recover any entry $P[i,\ell]$ or $Q[m,k]$, nor
    \item compute $P[i,\ell]\cdot Q[m,k]\bmod(p-1)$ beyond what $\Eval$ on held keys reveals.
\end{enumerate}
\noindent This is a standard indistinguishability assumption for two-input FE with index-restricted keys. The FE layer acts solely as a secrecy wrapper for exponents.

\subsection{RDMPF Algebra and Interface}\label{sec:formal-rdmpf}

We work over $\Zp^{\dim\times\dim}$ with public parameters $\pi_{\text{pub}}=\{p,\dim,\text{BaseX},\text{BaseY},W\}$, where
$\mathrm{rank}(\text{BaseX})=\mathrm{rank}(\text{BaseY})=\dim-1$ and $\mathrm{rank}(W)=\dim$.

\noindent Public matrices are formed by scalar actions on fixed bases
\[
P=\lambda\cdot\text{BaseX}\bmod(p-1),\qquad Q=\omega\cdot\text{BaseY}\bmod(p-1)
\]

\noindent for secret exponents $\lambda,\omega\in_R\Zpm$. By Lemma~\ref{lem:scalar_comm}, scalar multiples of the same base commute.

\paragraph{Non-interactive shared secret (sender $S$ to recipient $R$)}
\noindent Let the registry expose the encodings 
\[
\widehat P_S=\EncL(\mathsf{pp},P_S)\quad \text{and}\quad \widehat Q_S=\EncR(\mathsf{pp},Q_S)
\]

\noindent for the sender’s \emph{ephemeral} key and $\widehat P_R,\widehat Q_R$ for the \emph{recipient} key. The sender, using his per-user FE keys, computes 
for each output coordinate $(i,j)$
\[
T_1[i,j] := \rdmpf(P_S,W,Q_R)_{ij} = \prod_{\ell=1}^{\dim}\ \prod_{m=1}^{\dim} W_{\ell m}^{\ e_{i\ell,mj}}\ \bmod p
\]

\noindent where
\[
e_{i\ell,mj} := \Eval\big(\mathsf{pp},\sk^{\textsf{left}}_{(i,\ell),(m,j)};\ \widehat P_S,\widehat Q_R\big) = P_S[i,\ell]\cdot Q_R[m,j]\bmod(p-1)
\]

\noindent Symmetrically, using the FE keys for the \emph{right} input fixed to $\widehat Q_S$, the sender computes
\[
T_2[i,j] := \rdmpf(P_R,W,Q_S)_{ij} = \prod_{\ell=1}^{\dim}\ \prod_{m=1}^{\dim} W_{\ell m}^{\ f_{\ell i,mj}}\ \bmod p
\]

\noindent and
\[
f_{\ell i,mj} := \Eval\big(\mathsf{pp},\sk^{\textsf{right}}_{(\ell,i),(m,j)};\ \widehat P_R,\widehat Q_S\big) = P_R[\ell,i]\cdot Q_S[m,j]\bmod(p-1)
\]

\noindent Finally, the session matrix is formed using the composition operator
\[
\keyd{S}{R} := T_1\triangleright T_2
\]

\noindent The receiver computes the same $T_1,T_2$ (and $\keyd{S}{R}$) using his per-user FE keys with roles swapped. Since the underlying $\rdmpf$ and 
$\triangleright$ are unchanged, Theorem~\ref{thm:nika} continues to give $\keyd{S}{R}=\keyd{R}{S}$ verbatim.

\newpage
\vspace*{1em}
\paragraph{What outsiders learn?}
A third party, with only $(\mathsf{pp},\widehat P_S,\widehat Q_S,\widehat P_R,\widehat Q_R,W)$ and \emph{no} per-user FE keys, cannot recover any $P$/$Q$ entries 
or evaluate any $P\cdot Q$ products, hence cannot form the exponents needed by $\rdmpf$ and cannot derive $\keyd{S}{R}$.

\paragraph{\underline{KEM derivation and capsule (formal restatement)}}
(For reference, we restate the constructions from Section~\ref{sec:phase2}.)

Let $\enc(\cdot)$ be a fixed public byte encoding of matrices (row-major, elements reduced to $[0,p\!-\!1]$ with fixed-width little-endian). Set
\[
\begin{aligned}
    Z &:= \SHA\!\big(\enc(\keyd{S}{R})\big)\\[2pt]
    (\Kenc,\Kauth) &\leftarrow \HKDF\!\big(Z;\ \textsf{info}=\text{"rdmpf-kem"},\ \mathsf{salt}=\mathsf{nonce}\big)\\[2pt]
    \mathsf{tag} &:= \HMAC_{\Kauth}(\ctx)\\[2pt]
    \ctkem &:= (epk_S,\mathsf{nonce},\mathsf{tag})\\[2pt]
    \Hint &:= \SHA(\ctkem)\\[2pt]
    \Inner &:= \AEAD_{\Kenc}(\textsf{payload})
\end{aligned}
\]
\noindent Only the tuple $\{\Hint,\ctkem,\Inner\}$ is written by $I_1$ to storage canisters $C_i$ and later fetched by $I_2$. Neither sender nor receiver ever 
calls $C_i$ directly. Authorization at $I_2$ requires $h=\SHA(K_{\mathrm{auth}})$, verified against the value embedded in $\Inner$.

\subsection{Hardness Assumptions}\label{sec:formal-assumptions}

We state standard-style assumptions tailored to RDMPF in this setting.

\paragraph{\underline{Computational RDMPF (cRDMPF)}}

Given $\pi_{\text{pub}}$ and public $P=\lambda\cdot \text{BaseX}$, $Q=\omega\cdot \text{BaseY}$, computing $\rdmpf(P,W,Q)$ without $(\lambda,\omega)$ is infeasible.

\paragraph{\underline{Decisional RDMPF (dRDMPF)}}

Given $(\pi_{\text{pub}},P,Q)$, distinguishing $\rdmpf(P,W,Q)$ from uniform in $\Zp^{\dim\times\dim}$ is infeasible.\\

\noindent Note: dRDMPF implies cRDMPF. The IND-CPA proof (Theorem~\ref{thm:rdmpf-aead}) invokes cRDMPF. Hence dRDMPF is stated for completeness and may be useful 
for stronger security notions.

\paragraph{\underline{Extractor suitability of $Z$}}

For independently sampled sender/recipient keys, $\keyd{S}{R}=T_1\triangleright T_2$ has high conditional min-entropy so that
\[
(\Kenc,\Kauth)\leftarrow \HKDF(Z;\ \textsf{info}=\text{``rdmpf-kem''},\ \mathsf{salt}=\mathsf{nonce})
\]
\noindent is indistinguishable from random given all public data.

\subsection{Non-Interactive Encapsulation Objects}\label{sec:formal-objects}

From Section~\ref{sec:phase2} (non-interactive KA), the concrete objects are
\[
\begin{aligned}
    Z &\leftarrow \SHA\!\big(\enc(\keyd{S}{R})\big)\\[2pt]
    (\Kenc,\Kauth) &\leftarrow \HKDF\!\big(Z;\ \textsf{info}=\text{"rdmpf-kem"},\ \mathsf{salt}=\mathsf{nonce}\big)\\[2pt]
    \mathsf{tag} &\leftarrow \HMAC_{\Kauth}(\ctx)\\[2pt]
    \ctkem&=(epk_S,\mathsf{nonce},\mathsf{tag})\\[2pt]
    \Hint&=\SHA(\ctkem)\\[2pt]
    \Inner&=\AEAD_{\Kenc}(\textsf{payload})
\end{aligned}
\]

\subsection{Security Implications}\label{sec:formal-conseq}

To avoid duplication, we restate exactly what the protocol needs and defer to the theorems where they are proved.

\begin{itemize}[leftmargin=1.6em]
    \item \textbf{Correctness of KA} (Theorem~\ref{thm:correct})
    \item \textbf{IND-CPA for $\Inner$} (Theorem~\ref{thm:rdmpf-aead})
    \item \textbf{Authorization soundness at $I_2$} (Theorem~\ref{thm:auth})
    \item \textbf{Recipient-private discovery} (Theorem~\ref{thm:hint})
    \item \textbf{Reclaim (Alice-only)} (Theorem~\ref{thm:reclaim})
\end{itemize}

\subsection{Message and State Types (for reference)}\label{sec:formal-types}

\paragraph{\underline{Stored tuple at $C_i$}}
\[
\Tuple := \{\Hint,\ctkem,\Inner\}\quad \text{(optionally }+\ \mathsf{reclaim\_tag}\text{)}
\]

\paragraph{\underline{Noticeboard records}}
\[
\Announce(\mathsf{idx},\Hint,\mathsf{rendezvous\_token},\mathsf{code\_hash}(I_1))\quad \text{and}\quad \Finalize(\mathsf{idx},\{\DestructProof\})
\]

\paragraph{\underline{Witness attestations}}
\[
\Commit(\Hint,\Id(C_i)),\quad \DestructIntent(\Id(X),\mathsf{deadline})\quad \text{and}\quad \DestructProof(\Id(X))
\]
for $X\in\{I_1,I_2,C_i\}$.


\section{Complexity Analysis}

\subsection{Computational Complexity}

\subsubsection{RDMPF Operation}

For $X, W, Y\in\Zp^{\dim\times\dim}$, each output entry is (Definition~\ref{def:rdmpf})
\[
\rdmpf(X,W,Y)_{jk}
 \;=\;
 \prod_{\ell=1}^{\dim}\ \prod_{m=1}^{\dim}
   W_{\ell m}^{\,X_{j\ell}\,Y_{mk}}
 \bmod p.
\]

\newpage

\noindent \textbf{Per output element $(j,k)$} we have that
\begin{itemize}[leftmargin=1.6em]
    \item $\dim^2$ multiplications in $\Z_{p-1}$ to form exponents $e_{\ell m}=X_{j\ell}Y_{mk}\bmod(p-1)$
    \item $\dim^2$ modular exponentiations $W_{\ell m}^{\,e_{\ell m}}\bmod p$
    \item $\dim^2-1$ modular multiplications in $\Z_{p}$ to multiply the terms
\end{itemize}

\noindent For the \textbf{whole matrix} there are $\dim^2$ outputs, hence
\begin{itemize}
    \item Modular exponentiations $= \dim^4$
    \item Modular multiplies for products $= \dim^4-\dim^2$
\end{itemize}
\noindent Writing $\text{MExp}(p)$ for the cost of one modular exponentiation at modulus $p$
\[
\text{Time}=\Theta\!\big(\dim^4\cdot \text{MExp}(p)\big).
\]
\noindent Exponents are reduced modulo $(p-1)$, so their bitlength is $O(\log p)$. In practice, precomputation and reuse across rows/columns can reduce 
effective scaling to about $O(\dim^3)$, but the $\dim^4$ exponentiation count is inherent to the definition.

\subsubsection{Protocol Totals (RDMPF counts)}

Let $n$ be the number of ephemeral canisters. Then,
\begin{itemize}[leftmargin=1.6em]
\item \textbf{Per canister token} each $C_i$ computes one token $\rdmpf(P_i,W,Q_i) \Rightarrow n$ RDMPFs (at canisters)
\item \textbf{Alice--Bob shared key (non-interactive)}
  \begin{itemize}
  \item Using the nested form, if $\text{Token}_A=\rdmpf(P_S,W,Q_S)$ and $\text{Token}_B=\rdmpf(P_R,W,Q_R)$ are already available, each side computes exactly 
  one RDMPF to combine with the other’s token (across Alice+Bob this is $\mathbf{2}$ RDMPFs).
  \item If each party recomputes both $(T_1,T_2)$ independently, count $\mathbf{4}$
  \end{itemize}

\item \textbf{Per-canister transport sessions (unlinkability)}
  \begin{itemize}
    \item Alice and Bob each compute one RDMPF per contacted $C_i$ using $\text{Token}_{C_i}$
    \item Total $\mathbf{2n}$ RDMPFs across both principals
  \end{itemize}
\end{itemize}

\noindent\textbf{Summary}
\[
\boxed{\text{Total RDMPFs} \in \{\,3n+2,\ 3n+4\,\}}
\]
depending on whether $(T_1,T_2)$ are recomputed or reused via tokens. Symmetric-crypto costs (AEAD/HMAC) are negligible compared to RDMPF.

\subsection{Communication Complexity}

Each token or shared key is a $\dim\times\dim$ matrix over $\Z_p$
\[
\text{size} \;=\; \dim^2 \cdot \lceil \log_2 p \rceil/8 \ \text{ bytes}
\]
For $p\approx 2^{192}$ this is $\dim^2\cdot 24$ bytes. Example: $\dim=12\Rightarrow 144\times 24=3456$ bytes $\approx 3.4\,\mathrm{KB}$.

\begin{table}[h]
\centering
\begin{tabular}{@{}lc@{}}
\toprule
Component & Size (typical) \\ \midrule
Token / key matrix & $\dim^2\cdot 24$ bytes, or $\approx 3.4\,\mathrm{KB}$ at $\dim=12$ \\
Outer package $\Inner$ (sealed payload) & $|M|$ + $\text{AEAD overhead}$ $(\approx 1\!-\!2\,\mathrm{KB})$ \\
\midrule
Per canister on-wire & $\approx 5\!-\!6\,\mathrm{KB}$ \\
Total for $n=2$ & $\approx 10\!-\!12\,\mathrm{KB}$ \\
\bottomrule
\end{tabular}
\caption{Communication costs. If tokens are computed in-place by $C_i$, on-wire is just $\Inner$.}
\end{table}

\subsection{Storage Complexity}

Each $C_i$ persists only the sealed tuple and minimal metadata until retrieval and destruction.
\begin{itemize}
  \item Stored tuple: $\Tuple = \{\Hint,\ctkem,\Inner\}$ ($\approx$ 2--10 KB depending on $\dim$ and payload header).
  \item Runtime state: transient request/response buffers (not persisted).
  \item \textbf{No RDMPF keys or decapsulation state} are stored on any canister at rest.
\end{itemize}


\section{Conclusion}

This paper presents a non-interactive key-agreement mechanism (RDMPF) paired with a ephemeral execution architecture that forwards only sealed payloads, certifies 
minimal public state (notice and destruction proofs), and then reliably destructs all ephemeral instances. Cryptographically, the design yields sender anonymity, content 
privacy, and session unlinkability from the hardness of RDMPF and the extractor properties of HKDF. Architecturally, the absence of long-lived intermediaries gives 
strong forward secrecy. 

Unlike SNARK-based mixers, our protocol does not require a trusted setup or heavy proving; and unlike UTXO mixers, there is no global pool to correlate. The result is 
a practical blueprint for private transfers that composes rigorous algebra with an execution model designed to leave nothing sensitive behind. The cryptographic core 
derives a shared secret as $\keyd{S}{R}$ from which $(\Kenc, \Kauth)$ are extracted to seal content and authenticate retrieval, respectively. On the systems side, the short-lived intermediaries 
$(I_1, I_2, C_i, \W)$ handle storage and routing in a strictly sealed manner. Canister $I_1$ accepts the tuple ${\Hint,\ctkem,\Inner}$, writes 
an $\Announce$ on $\Notice$, and distributes the sealed payload to $C_i$. Each $C_i$ canister stores only that tuple and serves it once to 
$I_2$ upon authorized retrieval. The witness $\W$ records $\Commit$, $\DestructIntent$, and $\DestructProof$ to certify what happened and when. Controllers are 
blackholed, and their parent (Factory) destroys them within TTL upon delivery by $I_2$ so no long-term keys are ever installed and no decryptable transcripts 
persist. After teardown, the only remnants are certified, low-leakage public records $(\Announce$, $\Finalize)$ and destruction proofs.

Under the stated assumptions (computational/decisional hardness for RDMPF, extractor/PRF security of $\HKDF$, PRF security of $\HMAC$, and IND-CPA of the chosen 
$\AEAD$) our design yields sender anonymity, content privacy from intermediaries, and session unlinkability. The ephemeral infrastructure and certified destruction 
provide strong forward-secrecy properties even against later cryptanalysis. From an implementation perspective, it is compatible with the ICP’s model (in terms of 
certified state and canister lifecycle). 

Overall, this protocol offers a practical, testable path to transaction privacy with auditable system invariants and minimal trust in infrastructure longevity.\\

\vspace{1em}
\noindent{\cogent{In recognition and appreciation to an exceptional colleague and scholar, Hugo Daniel Scolnik. A towering figure in Latin American computer science, 
operational research and cybersecurity, I am grateful for his friendship, academic cooperation, and the countless conversations that influenced much of the work leading 
to this paper.}}



\newpage
\appendix

\section{Notation}
For convenience to the reader, we provide below the notation that has been used throughout the paper.

\begin{table}[h]
\centering
\begin{tabular}{@{}ll@{}}
\toprule
Symbol & Meaning \\ \midrule
$p$ & Prime modulus (at least 192 bits) \\
$\Zp$ & Integers modulo $p$ \\
$\Zpm$ & Integers modulo $p-1$ (for exponents) \\
$\dim$ & Matrix dimension ($\approx 8$--$24$) \\
$\in_R$ & Chosen uniformly at random \\
$\|$ & Concatenation operator \\
$[\![n]\!]$ & Set $\{1,2,\ldots,n\}$ \\[0.25em]
\midrule
$epk_R=(P_R,Q_R)$ & Receiver's (Bob) public parameters \\
$epk_S=(P_S,Q_S)$ & Sender's (Alice) one-time public key (per transfer) \\
$\mathrm{RDMPF\_Key}((P_S,Q_S),(P_R,Q_R))$ & Non-interactive RDMPF key derivation \\
$Z$ & RDMPF shared secret output \\
$(\Kenc,\Kauth)$ & Transport keys via $\HKDF(Z;\textsf{info}=\text{"rdmpf-kem"}$,\\ & $\textsf{salt}=\mathsf{nonce})$ \\
$\mathsf{nonce}$ & HKDF salt (e.g., 256-bit) \\
$\ctx$ & Bound context (Factory tx id, canister ids, \\ & subnets, TTL, code hash, params) \\
$\mathsf{tag}$ & $\HMAC_{\Kauth}(\ctx)$ \\
$\ctkem$ & Capsule $(epk_S \| \mathsf{nonce} \| \mathsf{tag})$ \\
$\Hint$ & Public discovery hint $\Hint=\SHA(\ctkem)$ \\
$\Inner$ & Inner ciphertext $\AEAD_{\Kenc}(\textsf{payload})$ \\
$\IV$ & AEAD nonce/IV (random per $\Kenc$) \\
$h$ & Authorization hash $\SHA(\Kauth)$ embedded in $\Inner$ \\[0.25em]
\midrule
$I_1, I_2$ & Deposit and retrieval intermediaries (ephemeral) \\
$C_i$ & Storage canisters (sealed, ephemeral), $i\in[\![n]\!]$ \\
$\W$ & Witness canister (ephemeral) \\
$\Notice$ & Certified Noticeboard (permanent) \\
$t, n$ & Quorum threshold $t$ and number of storage canisters $n$\footnote{The production deployment uses $n=2$ storage canisters to minimize spawn costs. 
The protocol generalizes to arbitrary $n$ with threshold $t \le n$.} \\[0.25em]
\midrule
$\rdmpf(X,W,Y)$ & Rank-Deficient Matrix Power Function (family name) \\
$\HKDF$ & HMAC-based key derivation (RFC 5869) \\
$\HMAC$ & HMAC-SHA3-256 \\
$\SHA$ & SHA3-256 (FIPS 202) \\
$\mathsf{BLAKE2s}$ & Domain-separated digests (witness events, context commits) \\
$\AEAD$ & ChaCha20-Poly1305 (RFC 8439) \\
$\Trunc_k(x)$ & Truncation of $x$ to $k$ bits \\ 
\bottomrule
\end{tabular}
\caption{Notation Summary}
\end{table}

\end{document}